\documentclass[12pt]{article}
\pdfoutput=1
\usepackage[centertags]{amsmath}
\usepackage{amsfonts}
\usepackage{amssymb}
\usepackage{amsthm}
\usepackage{graphicx}

\newcommand{\beq}{\begin{equation}}
\newcommand{\eeq}[1]{\label{#1}\end{equation}}
\newcommand{\bea}{\begin{eqnarray}}
\newcommand{\eea}[1]{\label{#1}\end{eqnarray}}
\newcommand{\p}{\partial}

\begin{document}
\setlength{\topmargin}{-1cm} \setlength{\oddsidemargin}{0cm}
\setlength{\evensidemargin}{0cm}

\begin{titlepage}
\begin{center}
{\Large \bf Fermion Dipole Moment and Holography}

\vspace{20pt}

{\large Manuela Kulaxizi$^{a, b}$ and Rakibur Rahman$^c$}

\vspace{12pt}
$^a$ School of Mathematics, Trinity College Dublin, Dublin 2, Ireland\\
\vspace{6pt}
$^b$ Lorentz Institute for Theoretical Physics, Leiden University\\
     P.O. Box 9506, Leiden 2300RA, The Netherlands\\
\vspace{6pt}
$^c$ Max-Planck-Institut f\"ur Gravitationsphysik (Albert-Einstein-Institut)\\
     Am M\"uhlenberg 1, D-14476 Potsdam-Golm, Germany

\end{center}
\vspace{20pt}

\begin{abstract}
In the background of a charged AdS black hole, we consider a Dirac particle endowed with an arbitrary magnetic
dipole moment. For non-zero charge and dipole coupling of the bulk fermion, we find that the dual boundary theory
can be plagued with superluminal modes. Requiring consistency of the dual CFT amounts to constraining the strength
of the dipole coupling by an upper bound. We briefly discuss the implications of our results for the physics of
holographic non-Fermi liquids.
\end{abstract}

\end{titlepage}

\newpage

%%%%%%%%%%%%%%%%%%%%%%%%%%%%%%%%%%%%%%%
\section{Introduction}\label{sec:Intro}
%%%%%%%%%%%%%%%%%%%%%%%%%%%%%%%%%%%%%%%

Gauge/gravity duality~\cite{Maldacena,Gubser,Witten} may serve as a diagnostic tool for assessing the consistency of a theory
since it allows one to explore otherwise unaccessible regions of the parameter space. Some pathological features of the theory
may become more manifest in one side of the duality or the other. Notable examples include the observed violation of causality
in the boundary CFT for some apparently healthy bulk duals with generic couplings: ghost-free Gauss-Bonnet gravity in five
dimensions~\cite{Myers,Buchel,Hofman}, and Lovelock and other higher-derivative theories of gravity~\cite{GB,Lovelock}.
The bulk couplings are constrained by the consistency of the boundary theory, and in the context of field theory this can be
understood as constrains imposed by unitarity~\cite{Hofman,Hofman1,Kulaxizi}. It was not until recently~\cite{Camanho} that
these bulk gravity theories are shown to have unambiguous signs of pathologies at the classical level.

The techniques of Ref.~\cite{Camanho} however cannot be employed for electromagnetic (EM) couplings\footnote{For EM couplings,
the scattering amplitude goes like the Mandelstam variable $s$, instead of $s^2$. This gives zero time delay for a fast particle
crossing a shock wave. We thank A.~Zhiboedov for this point.}. To date, there is no known classical inconsistency for a massive
charged spin-1 particle to have an arbitrary magnetic dipole moment~\cite{VZ}; only quantum consistency requirements, like
perturbative renormalizability and tree level unitarity, spell out a unique bare value at the Lagrangian level~\cite{Tiktopoulos}.
Yet it was argued in Ref.~\cite{KR} that once the theory is placed in a charged AdS black hole background, a generic dipole
coupling would afflict the dual CFT with superluminal modes. A similar story is expected for the magnetic dipole coupling of a
massive Dirac fermion, which we are going to consider in this article.

Note that if the EM interactions preserve Lorentz, parity and time-reversal symmetries, a spin-$\tfrac{1}{2}$ particle
of mass $m$ may possess a charge $q$ and a magnetic dipole moment $\mu_m$. The Lagrangian for a Dirac field $\Psi$
incorporates them through the minimal coupling and the Pauli term:
\beq \mathcal{L}=-i\bar{\Psi}(\displaystyle\gamma^\mu D_\mu-m)\Psi+q\lambda\bar{\Psi}\,
\gamma^{\mu\nu}F_{\mu\nu}\Psi,\eeq{L00}
where the covariant derivative is $D_\mu=\partial_\mu-iqA_\mu$, and the $\Gamma$ matrices are the same as those of Ref.~\cite{Guarrera}. 
The magnetic dipole moment has the value
\beq \mu_m=\frac{q}{2m}\left(2+4m\lambda\right).\eeq{dipoles}
The minimal coupling alone gives a gyromagnetic ratio of 2, as shown by Dirac. But nothing forbids a non-zero contribution from
the dipole coupling since the classical electrodynamics is consistent for an arbitrary value of $\lambda$. However, the Pauli
term renders the theory power-counting non-renormalizable, and so quantum consistency requires that $\lambda$ vanish at the
tree level~\cite{Weinberg}. In an effective field theory, though, quantum corrections would generate a small but non-zero
value for this coupling constant.

When coupled to gravity as well, the system would be described by the Lagrangian:
\beq \mathcal{L}=-i\sqrt{-\det g}\left[\,\bar{\Psi}\left(\Gamma^\mu\nabla_\mu-m\right)\Psi+iq\lambda\bar{\Psi}\,
\Gamma^{\mu\nu}F_{\mu\nu}\Psi\,\right],\eeq{L0}
where $\Gamma^\mu=e^\mu_a\gamma^a$ is given by the flat-space $\gamma$-matrices and the vierbein, the covariant derivative
$\nabla_\mu=D_\mu+\tfrac{1}{4}\hat{\omega}_{\mu ab}\gamma^{ab}$ incorporates the spin connection, and
$\Gamma^{\mu\nu}=e^\mu_ae^\nu_b\gamma^{ab}$.
%The non-minimal coupling shows up naturally in top-down AdS/CFT
%models~\cite{Grana:2002tu,Ammon,CT,DeWolfeab,Gauntlett:2011mf,Gauntlett:2011wm}.

When the fermion is a probe in an AdS Reissner-Nordstr\"om background, the system~(\ref{L0}) has a boundary dual that captures
the physics of a non-Fermi liquid~\cite{Vegh}. The dipole coupling has been extensively studied in
bottom-up~\cite{Guarrera,Edalati,Li,Phillips} and top-down~\cite{Ammon,CT,DeWolfeab,Gauntlett} AdS/CFT models.
In this article we will use the same setup to show that holography poses a constraint on the bulk dipole coupling.

The organization of the article, with Section~\ref{sec:Main} constituting the bulk, is as follows. Given the AdS Reissner-Nordstr\"om
geometry, we consider in Section~\ref{sec:Fluctuations} the equations of motion of a probe fermion, which simplify considerably in
some suitable region of the parameter space, where the momentum, frequency, chemical potential and mass-squared are all very large.
The resulting zero-energy Schr\"odinger problem is considered in Section~\ref{sec:Constraints} to  show that there exist normalizable
solutions peaked near the boundary. The conditions that allow for such solutions also allow us to compute the group velocity of certain
modes coupled to the boundary fermionic operator. We derive an upper bound on the strength of the dipole coupling, which when violated
plagues the boundary theory with acausality. This result is reconfirmed in Section~\ref{sec:WKB}, where we use the WKB method and
compute numerically the group velocity of the boundary modes from the resulting Bohr-Sommerfeld quantization condition. In
Section~\ref{sec:Conclusions}, we make some remarks and briefly discuss the implications of our results for the physics of holographic
non-Fermi liquids.

%%%%%%%%%%%%%%%%%%%%%%%%%%%%%%%%%%%%%%%%%%%%%%%%%%%%%%%%%%%%%%%%%%%%%%%%%%
\section{Fermion Dipole Coupling \& Holographic Diagnosis}\label{sec:Main}
%%%%%%%%%%%%%%%%%%%%%%%%%%%%%%%%%%%%%%%%%%%%%%%%%%%%%%%%%%%%%%%%%%%%%%%%%%

The charged AdS black hole is a solution of the Einstein-Maxwell system in $\text{AdS}_{d+1}$~\cite{AdSRN}:
\beq ds^2=\frac{r^2}{L^2}\left[-f(r)dt^2+d\vec{x}^2\right]+\frac{L^2}{r^2}\frac{dr^2}{f(r)},\qquad A_\sigma=\mu\left(1-{r_0^{d-2}\over
r^{d-2}}\right)\delta_\sigma^t,\eeq{solution}
where $L$ is the AdS radius. The function $f(r)$ and $\mu$ generically depend on the mass and the charge of the black hole, but reduce
in the zero temperature limit of the black hole to
\beq f(r)=1+{d\over d-2}\left({r_0\over r}\right)^{2d-2}-{2(d-1)\over d-2}\left({r_0\over r}\right)^d,\qquad \mu=\pm\sqrt{\tfrac{1}
{2}d(d-1)}\,\frac{g_Fr_0}{(d-2)L^2}\,,\eeq{s6}
where $r_0$ is the horizon radius and $g_F$ is the effective dimensionless gauge coupling~\cite{Vegh}.

In this background, it is a great convenience that certain cubic and higher couplings
could be investigated at the level of a quadratic Lagrangian of the bulk field(s). From the CFT point of view, it means that 2-point
functions at a finite charge density contain information about certain 3- and higher-point functions.

%%%%%%%%%%%%%%%%%%%%%%%%%%%%%%%%%%%%%%%%%%%%%%%%%%%%%%%%%
\subsection{Fermion Fluctuations}\label{sec:Fluctuations}
%%%%%%%%%%%%%%%%%%%%%%%%%%%%%%%%%%%%%%%%%%%%%%%%%%%%%%%%%

Considering a probe Dirac fermion in the $(d+1)$-dimensional AdS Reissner-Nordstr\"om geometry, given by Eqs.~(\ref{solution}) and~(\ref{s6}),
allows us to study the dynamics of a dual fermionic operator $\mathcal O$  at zero temperature but at a finite chemical potential $\mu$. The mass $m$
of the field, taken to be positive without any loss of generality, is related to the conformal dimension $\Delta$ of the operator as:
$\Delta={d\over2}+mL$, assuming a stable CFT for small mass~\cite{Vegh}. The $d$-dimensional CFT has a global $U(1)$ symmetry and an associated
conserved current $J_\mu$, which the holographic duality maps to the bulk $U(1)$ gauge field. The CFT operator $\mathcal O$ has a charge
$q$ under the global $U(1)$.

In the following we restrict ourselves to $d=3$. For future convenience, let us choose the black hole charge $Q$ such that
$\text{sgn}(q)=\text{sgn}(Q)=\text{sgn}(\mu)$. Without any loss of generality, we take them all to be positive.
The dynamics of a probe spin-$1\over2$ field in the background is governed by
the Dirac equation obtained by varying of the action~(\ref{L0}):
\beq (\Gamma^\mu\nabla_\mu-m)\Psi+iq\lambda\,\Gamma^{\mu\nu}F_{\mu\nu}\Psi=0.\eeq{eom1}
Rotational invariance allows us to consider small perturbations of the form $\Psi(r,t,x)$, which can be Fourier transformed as
\beq  \Psi(r,t,x)=\left(-\det{\!g}\,g^{rr}\right)^{-\tfrac{1}{4}}\int \frac{d\omega d k}{(2\pi)^2}\,\psi(r,\omega,k)
\,e^{i(kx-\omega t)}.\eeq{ftW}
Substituting~(\ref{ftW}) into Eq.~(\ref{eom1}) one finds:
\beq \left(\gamma^r\partial_r-m\sqrt{g_{rr}}\right)\psi-i\left[\sqrt{\frac{g_{rr}}{-g_{tt}}}\gamma^t\left(\omega+q\mu
\left(1-\frac{r_0}{r}\right)\right)-\sqrt{\frac{g_{rr}}{g_{ii}}}\gamma^xk-\frac{2q\lambda\mu r_0}
{\sqrt{-g_{tt}}\,r^2}\,\gamma^{rt}\right]\psi=0.\eeq{eom2}
Let us introduce a new radial variable:
\beq \zeta\equiv \frac{r_0}{r},\qquad 0\leq\zeta\leq1.\eeq{rhodef}
Now the boundary is at $\zeta=0$ and the horizon at $\zeta=1$. We also define the parameters:
\beq \tilde{\omega}=\frac{\omega L^2}{r_0}\,,\qquad \tilde{k}=\frac{k L^2}{r_0}\,,\qquad \tilde{\mu}=\frac{q\mu L^2}{r_0}\,,\eeq{uvdef}
Denoting a derivative w.r.t. $\zeta$ by a prime, we rewrite the Dirac equation~(\ref{eom2}) as
\beq \psi^{\prime}+\mathcal{A}\psi=0,\eeq{eom3}
where the matrix-valued function $\mathcal{A}=\mathcal{A}(\zeta)$ is given in terms of the Pauli matrices $\sigma^{1,2,3}$:
\beq
\mathcal{A}=\left(
\begin{aligned}
\begin{tabular}{c  c }
  $\frac{-\tilde k+2\tilde\mu\zeta\lambda/L}{\sqrt{f(\zeta)}}\,\sigma^1+\frac{\Omega(\zeta)}{f(\zeta)}\,i\sigma^2-\frac{mL}
  {\zeta\sqrt{f(\zeta)}}\,\sigma^3$ & $\mathbf 0$\\\\$\mathbf 0$ & $\frac{\tilde k+2\tilde\mu\zeta\lambda/L}
  {\sqrt{f(\zeta)}}\,\sigma^1+\frac{\Omega(\zeta)}{f(\zeta)}\,i\sigma^2-\frac{mL}{\zeta\sqrt{f(\zeta)}}\,\sigma^3$\\
\end{tabular}
\end{aligned}
\right),
\eeq{Adef}
with
\beq f(\zeta)=1-4\zeta^3+3\zeta^4,\qquad\Omega(\zeta)=\tilde{\omega}+\tilde{\mu}(1-\zeta).\eeq{kappa}

For the sake of better physical understanding, let us apply the operator $\left(\mathbf{1}\partial_\zeta-\mathcal{A}\right)$ on
Eq.~(\ref{eom3}) to obtain a multi-channel coupled Schr\"odinger problem:
\beq -\psi^{\prime\prime}+\left(\mathcal{A}^2-\mathcal{A}^{\prime}\right)\psi=0.\eeq{eom5}
Because the matrix $\mathcal{A}$ is block diagonal, the 2-component chiral spinors comprising $\psi$ decouple from each other.
In other words, we can write
\beq \psi={\psi_-\choose\psi_+},\eeq{chirals}
and split Eq.~(\ref{eom5}) into two: one with $\psi_-$ alone and the other with $\psi_+$. However, the components
of $\psi_\pm$ themselves are non-trivially coupled, which makes the problem difficult.

We will now focus on some suitable region of the parameter space, where the momentum, frequency, chemical potential and mass-squared
are all very large. The system~(\ref{eom5}) will be much simplified in this regime. To spell out the region of parameter space to be
considered, let us first define some dimensionless ratios:
\beq u\equiv\frac{\tilde\omega}{\tilde{k}}\,,\qquad v\equiv\frac{\tilde{\mu}}{\tilde{k}}\,,\qquad w\equiv
\frac{m^2L^2}{\tilde{k}}\,.\eeq{uvwdefined}
Now we take the following limit of large momentum that simultaneously sets the frequency, chemical potential and mass-squared large
as well:
\beq \tilde{k}\gg1,\qquad\text{with}\quad u,\,v,\,w=\text{constant}.\eeq{limita}

This limits our focus at a small corner of the vast parameter space of the boundary field theory: one is choosing a large scaling
dimension for the operator $\mathcal O$, tuning the chemical potential as large as $\Delta^2$, and looking at comparable frequencies
and momenta.

Note that because of the large chemical potential the spinor field is subject to instability or condensation deep inside the
bulk where the background EM field is sufficiently strong. These effects are negligible as long as the EM field invariant is small:
$|F_{\mu\nu}|^2\ll\tfrac{m^4}{q^2}$. In the vicinity of the boundary, a region $[0,\,\zeta]$ does not encounter these issues as long
as $\zeta$ is small enough: \beq \left(\tfrac{v}{w}\right)^2\zeta^4\ll1.\eeq{Schwinger}
This means in particular that if $v,w\sim1$, as is chosen in Section~\ref{sec:Constraints}, the idea of a long-lived propagating particle
makes sense only in the near-boundary region, which is where one should focus at. As long as the condition~(\ref{Schwinger}) is satisfied,
other values of the parameters may render it safe to consider deep regions in the bulk as well (see Section~\ref{sec:WKB}).

%%%%%%%%%%%%%%%%%%%%%%%%%%%%%%%%%%%%%%%%%%%%%%%%%%%%%%%%%%%%%%%%%%%%%%
\subsection{Causality Violation \& Constraints}\label{sec:Constraints}
%%%%%%%%%%%%%%%%%%%%%%%%%%%%%%%%%%%%%%%%%%%%%%%%%%%%%%%%%%%%%%%%%%%%%%

Let us set the AdS radius to unity: $L=1$. In the limit~(\ref{limita}), the two components of $\psi_\pm$ decouple from each other,
except for the near-horizon region. The single-channel quantum mechanical problem one obtains in the limit~(\ref{limita}) is a \emph{zero-energy}
Schr\"odinger equation:
\beq -\frac{1}{\tilde k^2}\,\psi^{\prime\prime}_\pm+V_\pm(\zeta)\psi_\pm=0,\eeq{schrodinger}
where the index labeling the different components of $\psi_\pm$ has been omitted (since the equations are identical), and the
approximate potential function is given by
\beq V_\pm(\zeta)=\frac{w}{\tilde k\zeta^2f(\zeta)}+\frac{(1\pm2\lambda v\zeta)^2f(\zeta)-(u+v-v\zeta)^2}{f^2(\zeta)}
\,.\eeq{potential}
Despite a $\frac{1}{\tilde k}$-dependence, the first term in $V_\pm(\zeta)$ becomes increasingly important as one approaches
the boundary, and makes the potential go to $+\infty$ as $\zeta\rightarrow0$. On the other hand, deep inside the bulk $V_\pm(\zeta)$
has large negative values. Therefore, if the potential has a local minimum in the vicinity of the boundary, it will also have
a local maximum farther inside the bulk. Close to the boundary a pair of local minimum and maximum will always exist, either
for the $\psi_-$ or the $\psi_+$ modes, if the parameters obey
\bea u+v&=&2|\lambda|,\label{cond1}\\4\lambda^2&>&1.\eea{cond2}

To see this, let us take the near-boundary expansion of the derivative of the potential:
\beq V^\prime_\pm(\zeta)=-\frac{2w}{\tilde k\zeta^3}+(u+v\pm2\lambda)+2v^2\left(4\lambda^2-1\right)\zeta
-12\left(8\lambda^2-1\right)\zeta^2+\mathcal O\left(\zeta^3\right).\eeq{expand}
Depending on the sign of $\lambda$, the constant piece $(u+v\pm2\lambda)$ can always be set to zero either for the $\psi_-$ or the
$\psi_+$ modes. This gives the choice of parameters~(\ref{cond1}). Provided the condition~(\ref{cond2}) is also fulfilled,
$V^\prime_\pm(\zeta)$ has a zero at the point
\beq \zeta_{\text{min}}\approx\left[\frac{w}{\tilde kv^2\left(4\lambda^2-1\right)}\right]^{1\over4}.\eeq{min1}
For $v,w\sim1$, this is arbitrarily close to the boundary in the limit $\tilde k\rightarrow\infty$. That $\zeta_{\text{min}}$ is
in fact a minimum of the potential can be understood from its second derivative:
\beq V^{\prime\prime}_\pm(\zeta_{\text{min}})\approx8v^2\left(4\lambda^2-1\right)>0.\eeq{min2}
Furthermore, the choices~(\ref{cond1})--(\ref{cond2}) also ensure that there is maximum ``slightly'' away from the boundary,
where the $1\over\tilde k$-term is negligible; it is located at the point
\beq \zeta_{\text{max}}\approx\frac{v^2}{6}\left(\frac{4\lambda^2-1}{8\lambda^2-1}\right),\eeq{max}
which is at least an order of magnitude smaller than unity for $v\approx1$.

\newpage
\begin{figure}[ht]
\begin{minipage}[b]{0.5\linewidth}
\centering
\includegraphics[width=1\linewidth,height=.65\linewidth]{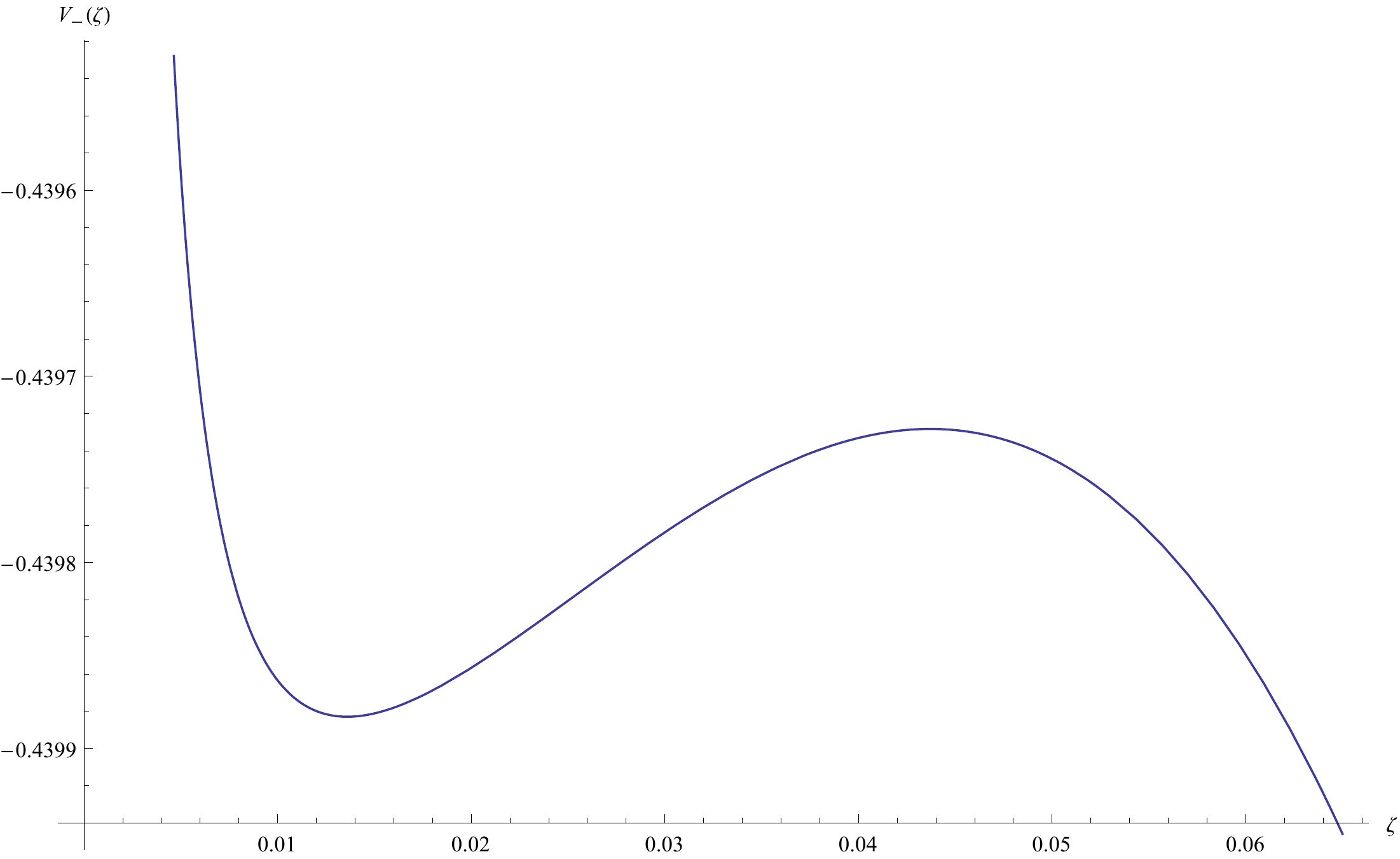}
\caption{\footnotesize $V_{-}(\zeta)$ is plotted for $\tilde k=10^8, v=1, w=1$ and $\lambda=0.6$, subject to the
conditions~(\ref{cond1})--(\ref{cond2}). The potential develops a well close to the boundary.
$V_{+}(\zeta)$ has an identical plot with the sign of $\lambda$ flipped.}
\label{fig:figure1}
\end{minipage}
\hspace{0.5cm}
\begin{minipage}[b]{0.5\linewidth}
\centering
\includegraphics[width=1\linewidth,height=.65\linewidth]{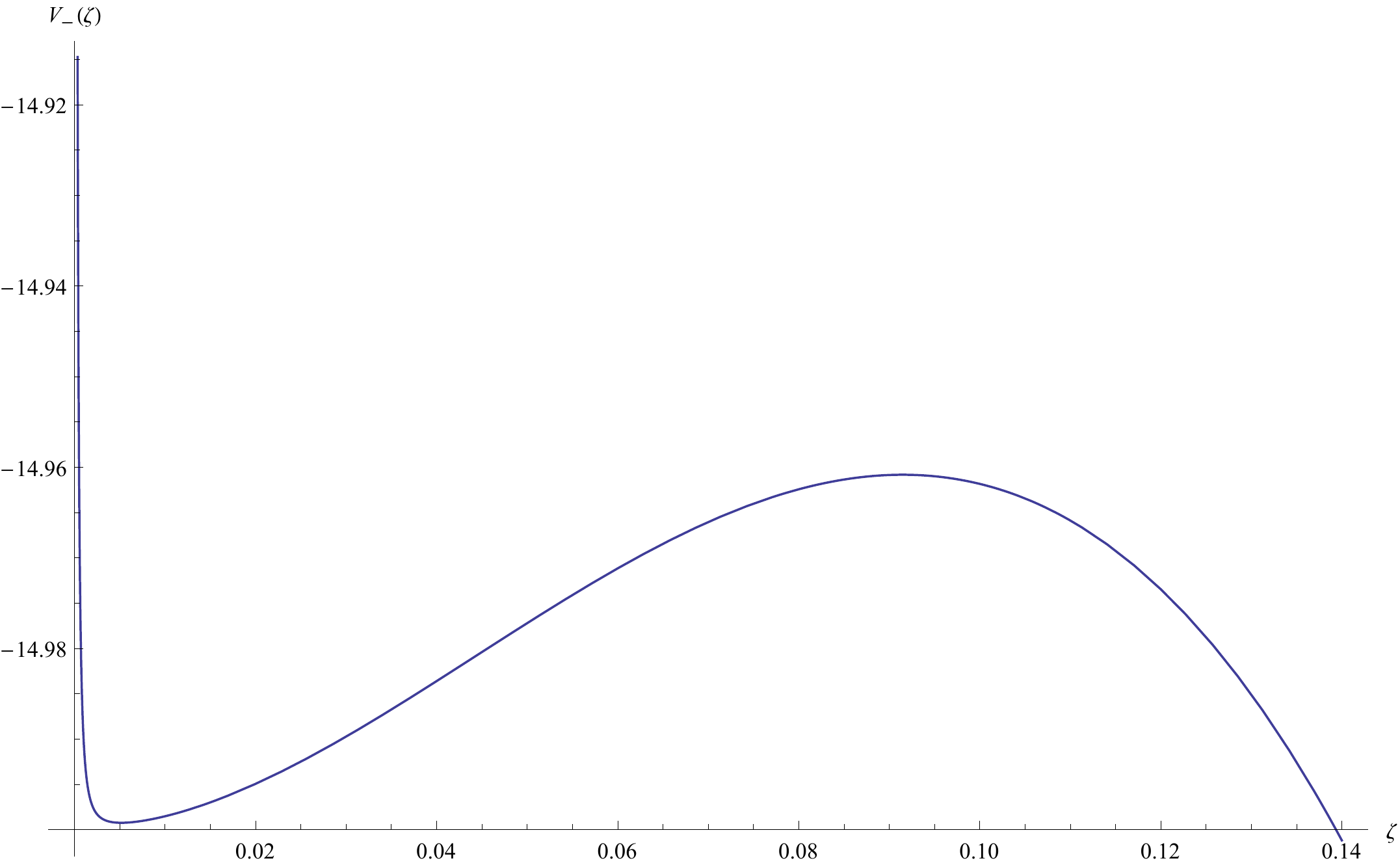}
\caption{\footnotesize $V_{-}(\zeta)$ is plotted for $\tilde k=10^8, v=1, w=1$ and $\lambda=2.0$, subject to the
conditions~(\ref{cond1})--(\ref{cond2}). The potential develops a well close to the boundary.
$V_{+}(\zeta)$ has an identical plot with the sign of $\lambda$ flipped.}
\label{fig:figure2}
\end{minipage}
\end{figure}

Under the conditions~(\ref{cond1})--(\ref{cond2}), therefore, the potential develops a well in the vicinity of the boundary,
where the EM field is sufficiently weak. Indeed, the inequality~(\ref{Schwinger}) holds very well for any
$\zeta\in[0,\zeta_{\text{max}}]$. The potential well will admit normalizable solutions of the zero-energy Schr\"odinger
problem~(\ref{schrodinger}) that are peaked near the boundary. For $\lambda>\tfrac{1}{2}$, the $\psi_-$ modes will have such
solutions, whereas if $\lambda<-\tfrac{1}{2}$, they occur for the $\psi_+$ modes.

The normalizable zero-energy solutions of the quantum mechanical problem correspond to the existence of (quasi)particles in
the dual CFT that are coupled to the operator $\mathcal O$. As one expects from Refs.~\cite{Myers,GB,Lovelock,KR}, the boundary
theory may suffer from causality violation for a generic dipole coupling. Such an inconsistency indeed arises
unless the bulk coupling parameter $\lambda$ is appropriately constrained, as we are going to see.
Note that the phase velocity of the modes coupled to $\mathcal O$ is given by
\beq c_p~\equiv~\frac{\tilde\omega}{\tilde k}~=~u.\eeq{cp0}
When this is compared with the condition~(\ref{cond1}), one gets a simple dispersion relation for the boundary modes, which takes the form:
\beq c_p~=~2|\lambda|-\frac{\tilde\mu}{\tilde k}\,.\eeq{up1}
This in turn allows one to compute the group velocity:
\beq c_g~\equiv~c_p+\tilde{k}\left(\frac{\partial c_p}{\partial\tilde k}\right)_{\tilde\mu}~=~2|\lambda|.\eeq{ug1}
Therefore, the group velocity exceeds unity on account of the condition~(\ref{cond2}). Because the boundary theory is
non-gravitational, this is an unambiguous signal of causality violation.

This problem can be avoided by requiring that the conditions~(\ref{cond1})--(\ref{cond2}) be never satisfied.
The only meaningful choice is to constrain the magnetic dipole coupling:
\beq \lambda^2\leq\frac{L^2}{4}\,,\eeq{constraint}
where the AdS radius $L$ has been restored. Outside this range of values, $-\tfrac{L}{2}\leq\lambda\leq+\tfrac{L}{2}$\,, the boundary CFT will
always be plagued with modes that propagate faster than light.

%%%%%%%%%%%%%%%%%%%%%%%%%%%%%%%%%%%%%%%%%%%%%%%%%%%%%%%%%%%%%%%%
\subsection{WKB Approximation: A Numerical Study}\label{sec:WKB}
%%%%%%%%%%%%%%%%%%%%%%%%%%%%%%%%%%%%%%%%%%%%%%%%%%%%%%%%%%%%%%%%

In this Section, we employ the WKB approximation method to reconfirm the occurrence of causality violation for $\lambda\notin\left[-\tfrac{L}{2},+\tfrac{L}{2}\right]$.
In the limit~(\ref{limita}), one can take $\hbar\equiv\tfrac{1}{\tilde{k}}$ and follow Refs.~\cite{Myers,Buchel,Hofman,GB,Lovelock} to write down a Bohr-Sommerfeld
quantization condition, which enables one to compute the group velocity of the dual field theory modes.

Let us recall from Eqs.~(\ref{eom5})--(\ref{chirals}) that while the chiral spinors $\psi_\pm$ are decoupled from each other, their components
themselves are not. Then, we start with the WKB ansatz:
\beq \psi_{\pm}^{\alpha}(\zeta)=\exp\left[{i\tilde{k}S_{\pm}^{\alpha}(\zeta)}\right],\qquad \alpha=1,2,\eeq{ans0}
where $\alpha$ labels the components of $\psi_\pm$, and $S_{\pm}^{\alpha}$ has an expansion in negative powers of $\tilde{k}$. In order for the ansatz~(\ref{ans0})
to satisfy Eq.~(\ref{eom5}) to all orders in $\tilde{k}$, it is required that $S_{\pm}^{\alpha}$ contain half-integer powers of $\tilde{k}$ as well.
Explicitly,
\beq S^\alpha_\pm\equiv S^\alpha_{0\pm}+\tilde{k}^{-1}S^\alpha_{1\pm}+\tilde{k}^{-{3\over 2}}S^\alpha_{{3\over 2}\pm}+\tilde{k}^{-2}S^\alpha_{2\pm}+\cdots\,.\eeq{ans1}

Substituting Eq.~(\ref{ans0})--(\ref{ans1}) into Eq.~(\ref{eom5}) one finds that the components of each chiral spinor also get decoupled and obey identical equations
to the leading order in ${1\over\tilde{k}}$. As a result, $S^\alpha_{0\pm}=S_{0\pm}$, i.e., the leading-order WKB phase is independent of the channel. The WKB
momentum, $p_\pm(\zeta)\equiv S^\prime_{0\pm}(\zeta)$, does not depend on the parameter $w$, and is given by
\beq p_\pm^2(\zeta)=\frac{(1\pm2v\zeta\lambda/L)^2f(\zeta)-(u+v-v\zeta)^2}{f^2(\zeta)}\,.\eeq{momentum}

To proceed, an additional assumption is required, namely $S^\alpha_{1\pm}=S_{1\pm}$. Note that the imaginary part of $S_{1\pm}$ gives the
WKB amplitude, while the real part the first-order phase correction, and again these are channel independent. Now one can go on with
the requirement that the determinant of the $2\times2$ coefficient matrix of $\psi_\pm$ vanish order by order in ${1\over\tilde{k}}$ (see, for example,
Refs.~\cite{wkb} for a discussion). At the next-to-leading order, the above requirement
results in an algebraic equation that determines $S^\prime_{1\pm}(\zeta)$, namely
\beq S^\prime_{1\pm}(\zeta)=i\partial_\zeta\left(\ln{\sqrt{p_\pm(\zeta)}}\right)+\text{Real Part}.\eeq{amplitude}
Clearly, the amplitude is inversely proportional to $\sqrt{p(\zeta)}$ like the single-channel case.

\newpage
Normalizability implies that $p_\pm^2(\zeta)<0$ in the boundary neighborhood. In the bulk there may exist turning points $\zeta_i$, where
the WKB momentum vanishes: $p_\pm(\zeta_i)=0$. For two turning points $\zeta_1$ and $\zeta_2$, one is lead to the Bohr-Sommerfeld quantization condition:
\beq \tilde{k}\int_{\zeta_1}^{\zeta_2}d\zeta\,p_\pm(\zeta)+\mathcal{O}(\tilde{k}^0)=\pi\left(n-\tfrac{1}{2}\right),\qquad \zeta_2>\zeta_1\quad
\text{and}\quad n=1,2,\dots\,.\eeq{qc}

From the holographic point of view, this can be understood as a dispersion relation for the boundary modes coupled to the dual fermionic operator.
Their group velocity $\left(\frac{\partial\tilde\omega}{\partial\tilde k}\right)_{\tilde\mu}$ can be computed by differentiating both sides of
Eq.~(\ref{qc}). The result is
\beq c_g^\pm=\frac{\int_{\zeta_1}^{\zeta_2}d\zeta\left(u \frac{\p p_\pm}{\p u} +v \frac{\p p_\pm}{\p v}-p_\pm\right)}{\int_{\zeta_1}^{\zeta_2}
d\zeta\left(\frac{\p p_\pm}{\p u}\right)}\,.\eeq{cg}

In what follows we resort to numerics to find some regions in the parameter space that allow for a pair of turning points to exist. This in turn enables
one to evaluate  numerically the group velocity from Eq.~(\ref{cg}). We take $\lambda>0$ and focus on the chiral spinor $\psi_-$ that serves our purpose
(for $\lambda<0$ we need to consider $\psi_+$ instead). For simplicity, we further restrict the exploration on the plane defined by Eq.~(\ref{cond1}) in the
3D space of $(u,v,\lambda)$.  The existence of turning points crucially depends on the values of the parameters: At $\lambda=\tfrac{L}{2}$, a single turning
point exists at $\zeta=\frac{1}{v}$. For $\lambda>\tfrac{L}{2}$, we distinguish two turning points $\zeta_1$ and $\zeta_2$, whose precise positions depend
on the parameter values. For fixed $v$, the distance $\zeta_2-\zeta_1$ grows with $\lambda$ as $\zeta_2$ moves towards the horizon. For fixed $\lambda$ and
increasing $v$, on the other hand, both turning points move closer to the boundary.

Figs.~3 and 4 show that the group velocity $c_g^{-}$ may indeed become superluminal for $\lambda>\tfrac{L}{2}$. For $\lambda$ slightly larger than this value,
causality violation is observed numerically, on the plane~(\ref{cond1}) in the parameter space, for $v>1$. As $\lambda$ increases, acausality may only
be captured by sufficiently large values of $v$. Note that the group velocity formula~(\ref{cg}) is independent of the parameter $w$, and so one can choose
the latter quantity large enough in order to accommodate large values of $v$ without violating the condition~(\ref{Schwinger}).

Thus Sections~\ref{sec:Constraints} and~\ref{sec:WKB} independently confirm our result~(\ref{cond1}). But they
differ in methodology and the region of parameter space taken into account. In  Section~\ref{sec:Constraints}, we took $v\approx1$ since
otherwise the maximum~(\ref{max}) would not lie close to the boundary, and this would invalidate the whole argument (recall that the single-channel
problem obtained in the limit~(\ref{limita}) would cease to make sense in the near-horizon region). Moreover, the condition~(\ref{cond1}), necessary for
the existence of a potential well near the boundary, was postulated to be the dispersion relation of the boundary modes. In contrast, in Section~\ref{sec:WKB}
the parameter $v$ was free to take large values, and causality violation was seen only for $v>1$ on the plane~(\ref{cond1}). More importantly, the boundary
dispersion relation~(\ref{qc}) was actually derived rather than postulated, which makes the analysis more rigorous.

\begin{figure}[ht]
\begin{minipage}[b]{0.5\linewidth}
\centering
\includegraphics[width=1\linewidth,height=.65\linewidth]{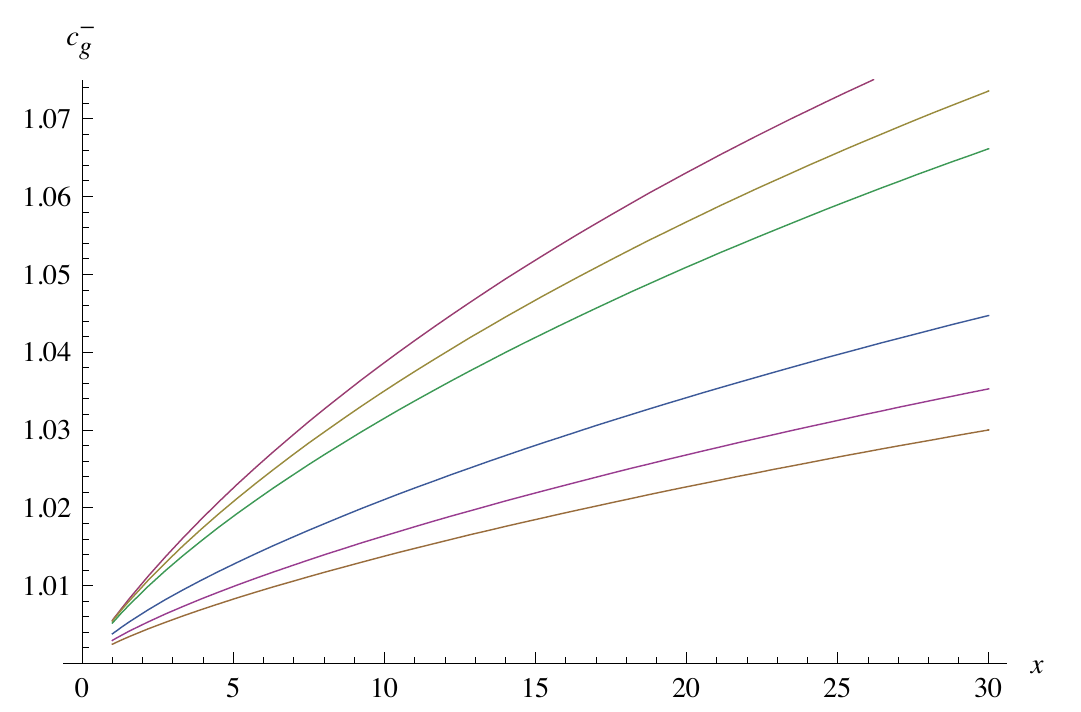}
\caption{\footnotesize $c_g^-$ is plotted against $x\equiv\left(\tfrac{\lambda}{L}-\tfrac{1}{2}\right)\times10^3$ for $(u,v)$ lying on the plane~(\ref{cond1}).
Different colors correspond to different values of $v=2,3,5,10,15,20$. The group velocity is found to exceed unity.}
\label{fig:figure3}
\end{minipage}
\hspace{0.5cm}
\begin{minipage}[b]{0.5\linewidth}
\centering
\includegraphics[width=1\linewidth,height=.65\linewidth]{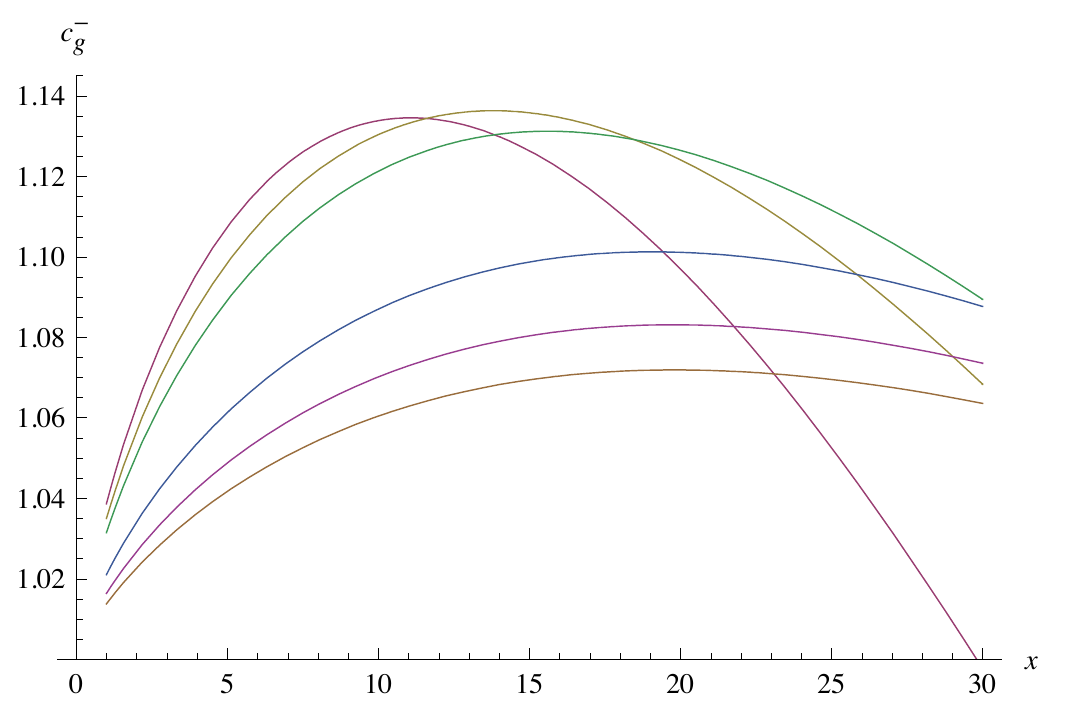}
\caption{\footnotesize $c_g^-$ is plotted against $x\equiv\left(\tfrac{\lambda}{L}-\tfrac{1}{2}\right)\times10^2$ for $(u,v)$  lying on the plane~(\ref{cond1}).
Different colors correspond to different values of $v=2,3,5,10,15,20$. The group velocity is found to exceed unity.}
\label{fig:figure4}
\end{minipage}
\end{figure}

\newpage
%%%%%%%%%%%%%%%%%%%%%%%%%%%%%%%%%%%%%%%%%%%%%%%%%%%%%%%
\section{Implications \& Remarks}\label{sec:Conclusions}
%%%%%%%%%%%%%%%%%%%%%%%%%%%%%%%%%%%%%%%%%%%%%%%%%%%%%%%

We have shown that the AdS/CFT correspondence poses non-trivial constraints on the EM interactions of a charged massive
Dirac particle: the otherwise undetermined strength of the dipole coupling of the classical bulk theory must have an
upper bound. We reduced the problem to a zero-energy Schr\"odinger equation and argued about the existence
of normalizable solutions peaked near the boundary when the parameters obey a certain relation. The latter served
as a dispersion relation for some corresponding boundary modes, which are found to propagate superluminally when the above
bound is violated. We also solved the coupled system of differential equations using the WKB approximation, and numerically
confirmed that the group velocity exceeds unity above this bound.

The bulk dipole coupling changes the structure of current-fermion-fermion 3-point functions of the CFT,
and one would like to investigate this point further. From the CFT point of view, constraints on $\lambda$
presumably correspond to constraints on the 4-point functions derived from unitarity and crossing symmetry. A promising
direction could go along the lines of Ref.~\cite{Komargodski}, where constraints related to charged operators were obtained.

Obtained by considering a probe fermion in the AdS Reissner-Nordstr\"om geometry, our results are directly relevant
for the physics of holographic non-Fermi liquids. Because large values of $\lambda^2$
are not physically meaningful, the disappearance of the Fermi surface and the onset of a gap, reported
in Ref.~\cite{Edalati} for $\lambda^2\simeq4L^2$, can actually never happen. Moreover, the shift in Fermi momentum
observed in~\cite{Guarrera} for varying $\lambda$ can only be small. In other words, not only is the existence
the Fermi surface robust but also the position thereof is sort of rigid for the allowed values of the dipole strength.
This fact is also supported by the study of several top-down models, e.g.~\cite{Gauntlett}.

A generic background may or may not have a holographic dual, but can only sharpen the bound~(\ref{constraint}).
On the other hand, one might argue that a more extensive holographic analysis covering other regions of the parameter
space could yield stronger constraints. In all the known examples~\cite{Myers,GB,Lovelock}, however, it has been
only the high momentum limit that revealed the CFT pathology. This may not come as a surprise since causality is likely
to be connected to the local, short-distance behavior of the theory~\cite{Myers}.

We have seen that a bulk theory with no known classical inconsistency may have a boundary dual that
exhibits pathologies. One is tempted to think that the CFT actually probes the quantum consistency of the theory
in the bulk, and that the duality holds good only if the latter is well behaved quantum mechanically, at least in the weak
coupling regime. An example that seems to justify this point is the existence of some classically consistent Vasiliev-like
theories in $\text{AdS}_3$ that are believed not to have healthy quantum versions since the dual CFTs are non-unitary~\cite{Perlmutter}.

%%%%%%%%%%%%%%%%%%%%%%%%%%%%%
\subsection*{Acknowledgments}
%%%%%%%%%%%%%%%%%%%%%%%%%%%%%

We would like to thank X.~O.~Camanho, T.~R.~Govindarajan, Y.~Korovin, A.~Parnachev, A.~O.~Starinets and A.~Zhiboedov for useful
discussions. MK would like to thank the organizers of the ``8th Crete Regional Meeting in String Theory,'' the ``9th QTS Meeting
in Yerevan'' and the Galileo Galilei Institute for Theoretical Physics for the hospitality, and the INFN for partial support
during the completion of this work. MK is supported in part by an NWO Vidi grant and by a Teaching Fellowship at Trinity College
Dublin.

\end{document}